\documentclass[10pt,conference]{IEEEtran}
\newtheorem{claim}{Claim}
\newtheorem{definition}{Definition}

\begin{document}

\title{Using Information Theory to Study the Efficiency and Capacity of Caching in the Computer Networks}

\author{
\authorblockN{Boris Ryabko}
\authorblockA{Siberian State University of Telecommunications and Information Sciences and \\Institute of Computational Technology of
Siberian Branch of Russian Academy of Science,  Novosibirsk, Russia\\
   boris@ryabko.net \\} 
   %\authorblockN{Nadezda Savina } % \authorblockA{Siberian State University of Telecommunications and Information Sciences,  Novosibirsk, Russia}
   }

\maketitle

\begin{abstract}
Nowadays  computer networks use different kind of memory whose speeds and capacities vary widely.
There exist   methods of a so-called caching which are intended to use the different kinds of memory in such a way that the frequently 
used data are stored in the faster memory, wheres the infrequent ones are stored in the slower memory.
We address the problems of estimating  the  caching   efficiency and its  
  capacity.   We define  the   efficiency and capacity of the caching   and   suggest
  a method for their estimation   based on the analysis of    kinds of the accessible  memory.
 %It is shown  how the suggested  method can be  applied to  estimate the performance of  caching.  
%In particular, this consideration gives a new look at the the organization of the cache memory.
%including a so-called cooperative caching.  
%Obtained results can be of some interest for  practical applications. 
\end{abstract}

{\em Keywords}  $\,$
network management, computer networks, future Internet design.

\section{Introduction} 

Nowadays computer networks are equipped with different kinds of memory whose capacity and speed are different and 
each user can use different types of memory   located in different network nodes.
That is why there are many special methods intended to use the different kinds of memory in such a way that the frequently 
used data are stored in the fast memory. As a rule, such methods are based on a so-called  caching (or cache memory).
There are a lot of papers devoted to methods of caching  and organizing the cache memory for  computers and 
networks, including such  practically important nets as content delivery networks;
see for review \cite{as,g,jcm,slb,spvd}. 
Note that we mainly say about  computer nets, but this consideration is valid 
and general networks which can contain all  mobile phones, satellite channels, data centers, etc. 
%Besides, our consideration is valid for  so-called Content Delivery Networks which are 

In this paper we suggest a new approach to investigate the performance of caching algorithms.
It is worth noting that close approach was used by the first author in \cite{Ry1} where the estimation of the computer performance was 
suggested.
In order to describe the main idea of the suggested approach we 
first shortly consider  a  content  distribution network.
Suppose there is a network with several nodes which store
some files (say, a set of movies $F$) and one node ($\omega$) can read files from others
and besides  times of reading are different from different nodes.  
Suppose, that if a needed file $f \in F$ is stored at several nodes, it is transmitted  from a node 
for which the time of transmission is  minimal and at one moment the node $\omega$ can read a file from one node.

Now some questions seem to be natural. Suppose,   this network  is used and 
    the frequency of  access of   files $f \in F$ obey  a certain distribution $p(f)$. 
    Either it is a good performance or it is far from the optimum? And what is the optimum?
    What are the maximal performance and the capacity?
There are several approaches to definitions of those 
concepts, see \cite{as,g,jcm,slb}. 

 We suggest the approach which gives a possibility to define the mentioned above notations and answer those questions
 basing on ideas of Information Theory. In order to describe the main idea we 
 first  give some definitions. Obviously, the main goal of the network is 
 to transmit some files (movies, songs, ets.) to their nodes (consumers). 
 That is why we call any sequence of files $f_1 f_2 ... f_n$, $n \ge 1$, 
  read by a node $\omega$ as   a  task  and denote a 
set of such sequences (tasks) which can be read by $\omega$ during a time $T$ as $N(T)$.

The key observation is as follows: for the large $T$, the number of tasks which can be carried out during
the time $2 T$  ($|N(2 T)|$) is (approximately)
$|N(T)|^2$, because the set of tasks (sequences of  files) $N(2 T)$
contains all concatenations of sequences  (tasks) carried out during the first time-interval  $T$ and the second one.
(Here and below $|S|$ is the number of elements of a set $S$.)
%Analogically,   the number of tasks which can be carried out during three hours ($N(3)$) is (approximately) $N^3$, etc. 
In other words, the number of  the file sequences which can be read 
during a certain time $t$ grows exponentially as a function of $t$, i.e. $|N(t)| = \exp (\alpha t)$, 
where $\alpha$ is a certain constant. 
Hence, it is natural to estimate an efficiency of the considered caching by the exponent $\alpha$
and this is the approach will be developed in this paper. Note that  this approach will be extended 
to case of estimation of the  throughput for general networks and cache memory of multi-core computers.

Let us briefly consider an example which can clarify the main idea of the approach. Let there be    two 
networks which have an identical  structure, but  their speeds of transmission    are different in such a way that the 
speed of the first net is  twice more than that of  the second one.  From  the   given
consideration we can see that the exponent $\alpha$ of the first net is twice more than the second one.
Apparently, it should be so.

This approach is  close, in spirit, to methods
of Information Theory, where, for example, the capacity of a  channel and the performance of an information source are
defined by the rate of  asymptotic growth    of the number of allowed sequences of basic symbols (letters),
(see \cite{co,S}). Besides, this approach was  recently applied to the definition of the efficiency and capacity 
of computers \cite{Ry1} which, in turn, found some practical applications \cite{fpr}.

\section{The capacity} 

In this part we define the capacity of a network equipped by a certain method of caching and suggest a 
 simple algorithm of calculation.
 Informally, the capacity equals the maximum 
throughput of the network and some simple examples show how the estimates of the capacity can be applied when, 
say, someone plans either to build 
or to modernize a network.

We consider a  network $\Omega$ formed by $n$ nodes $\omega_1, ..., \omega_n$. 
Suppose that the node $\omega_i$  stores a set of files $F_i$ and a node ($\omega$) can read files from some other nodes.
Let  $F$ is the set of all files, i.e. $F = \cup_{i=1}^n  F_i$.
%Also suppose that all files have the same size and d
If a needed file $f \in F$ is stored at several nodes, it is transmitted  from a node 
for which the time of transmission is  minimal and at one moment the node $\omega$ can read a file from one node.

Denote the time of reading a file $f$ by the node $\omega_i$ from the node
$\omega_j$ by $\tau_{ij}(f)$ and let $\tau_{\omega_i}(f)$ be the minimal time needed to obtain the file $f$ by the node
$\omega_i$, i.e.
\begin{equation}\label{taumin}
\tau_{\omega_i}(f) = \min_j  \tau_{ij}(f) \, \, .
\end{equation}
For the sake of simplification of notations, it will be convenient to let 
$\tau_{ij}(f) = \infty$, if $\omega_i$ cannot obtain $f$ from  $\omega_j$.
The following Fig 1 is an  example of probably the simplest network, where 
$\omega_1$ can be a server, $\omega_2$ can be the end user and the server cannot read files from the end user.
Here,  by definition, 
$\tau_{1 2 }(f) = \infty$.
\begin{figure}[h]\caption{The simplest network.}
\setlength{\unitlength}{0.75mm}
\begin{picture}(60,40)
\put(28,20){\circle{3}}
\put(30,20){\vector(1,0){30}}
\put(62,20){\circle{3}}
\put(65,20){$\omega_2$}
\put(21,20){$\omega_1$}
\end{picture}
\end{figure}
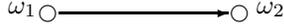
Let the node $\omega_i$ reads a sequence of files
 $f' = f_1, f_2, ... , f_s$.  We call any such a sequence as a task and define  the execution time $\tau_i(f)$ of the 
$f'$ by equation $$\tau_{\omega_i}(f') = \sum_{k=1}^s \tau_{\omega_i}(f_k) \, ,$$
i.e. it is supposed that any file $f_k$ is  transmitted at maximum speed. 
For a nonnegative $T$ we define as $\nu_i(T)$ the number of tasks of the node $\omega_i$,
whose execution 
 time equals $T$, i.e.
\begin{equation}\label{N}
\nu_i(T) =|  \{  f':  \tau_{\omega_i}(f') =  T \} |  \, .
\end{equation} 
We gave an informal explanation that $\nu_i$ grows exponentially, that is why  
we define the capacity of the node $\omega_i$ as follows:
\begin{equation}\label{capi}
C(\omega_i) \, = \, \limsup_{T \rightarrow\infty}  \frac{  \log \nu_i(T)} { T} \, 
\end{equation}
(bits per time unit) and let  the capacity of the network be the following sum of the node capacities
 \begin{equation}\label{cap} C(\Omega) \, = \, \sum_{\omega \in \Omega}  C(\omega) \, .
 \end{equation}
 (Here and below $\log x \equiv \log_2 x$.)
Of course, both capacities depend on the stored files and time of their transmissions.

%There are many natural situations where  $\lim$ can be used instead of $\limsup$ in (\ref{capi}). The following Claim describes  some of such cases.
%\begin{claim}\label{c1} The limit (\ref{capi}) exists if the execution times $\tau(f)$ are integers
%and the greatest common divisor of  $\tau(f),  x \in I,$ equals 1. \end{claim} {\it Proof} is given in Appendix, but here we note that $\lim$ can be used in all practically interesting situations. 

%Let us consider some simple examples in order to clarify what the defined capacity is.
The first question is how to estimate the capacity. 
The simple method for calculation $C(\omega)$ is well-known in Information Theory and was suggested 
by C.Shannon in 1948  \cite{S}, when he estimated the 
capacity of a lossless channel. Applying his method to the considered problems,
we can say that   the capacity of one node $C(\omega_i)$ is equal to the logarithm of the largest real 
solution $X_0$ of the following equation:
 \begin{equation}\label{X}
   \sum_{f \in F} X^{-\tau_{\omega_i}(f)}   = 1 \, . 
\end{equation} 
  In other words, $C(\omega_i) = \log X_0 . $ By definition, $C(\omega_i) = 0$, if this equation 
  does not have solutions (it is possible if all  $\tau_{\omega_i}(f) = \infty$).
Note that the root can be calculated  by a so-called bisection method
which is the simplest root-finding algorithm. 
As we mentioned above $C(\Omega) = \sum_{ \omega \in \Omega} C(\omega) $.

Let us consider two simple examples. First we look at the Fig. 1 and suppose that the first node 
$\omega_1$ is a library which stores $10^7$ files, wheres $\omega_2$ is an end user which can store $10$ files,
i.e. $|F_1| = 10^7$, $|F_2| = 10$. Also suppose that 
$\tau_{2 2 }(f) = 1$ for all $f \in F_2$  and $\tau_{1 2 }(f)  = 10$, for all $f \in F_1$,   
i.e. reading of a file from ``own'' memory ($F_2$) requires 1 time-unit, 
whereas reading from $\omega_1$ requires 10 time-units. Besides, the node $\omega_1$ cannot read files from $\omega_2$,
hence, $\tau_{2 1 }(f) = \tau_{2 2 }(f) = \infty $. 
Supposing that files in $F_1$ and $F_2$ are different ($F_1 \cap F_2 = 0$),  from the equation (\ref{X}) we obtain
\begin{equation}\label{ex}
 \frac {10} {X} + \frac{10^7}{ X^{10}} \, = \, 1 \, .
\end{equation}
Calculating, we obtain $X_0 = 10.01$, $C(\omega_2) = \log(10.01) = 3.324  $ bit per time-unit. 
Note that $C(\Omega) = 3.324  $, too, because $C(\omega_2)  = 0$. (Indeed, $\tau_{2 1 }(f) = \tau_{2 2 }(f) = \infty $, hence, 
 by definition, the capacity is 0.)

\begin{figure}[h]\caption{The three-node network.}
\setlength{\unitlength}{0.75mm}
\begin{picture}(60,40)
\put(28,20){\circle{3}}
\put(30,20){\vector(4,-1){30}}
\put(30,20){\vector(4,1){30}}
\put(62,28){\circle{3}}
\put(65,28){$\omega_2$}
\put(21,20){$\omega_1$}
\put(65,12){$\omega_3$}
\put(62,12){\circle{3}}
\put(62,13){\vector(0,1){13}}
\put(62,27){\vector(0,-1){13}}

\end{picture}
\end{figure}
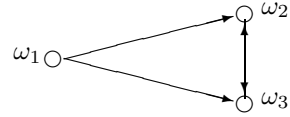

Now let us consider the  network on the Fig. 2. Suppose, that the nodes 
 $\omega_1$ and $\omega_2$ have the same parameters as in the previous example 
 and $\omega_3$ is an identical to $\omega_2$. Besides, let $\omega_2$ can read files from $\omega_3$,
 and, vise versa, $\omega_3$ can read files from $\omega_2$. Suppose that the time of both readings 
is two units per file, i.e. 
$\tau_{ 2 3  }(f) = \tau_{3 2 }(f) = 2$. 
If the set of files $F_1$ and $F_2$ are different ($F_3 \cap F_2 = 0$), we obtain for  the nodes
$\omega_2$ and $\omega_3$ the following identical  equations for calculating the capacities:
$$
 \frac {10} {X} + \frac {10} {X^2} + \frac{10^7}{ X^{10}} \, = \, 1 \, .
 $$
 Direct calculation shows that 
 $C(\omega_2) = C(\omega_3)   = 3.449  $ and, hence, the total capacity of the network $\Omega$ equals
 $ 2 \cdot 3.449 = 6.898$. 
 It is interesting that if the  nodes $\omega_2$ and $\omega_3$ store the same set of files ($F_2 = F_3$),
 the capacity of each of them will be equal the capacity of $\omega_2$ from the Fig.1.
 In other words, the capacity depends on content of the nodes. Naturally, 
 the speed of transmission, the memory size of different node and topology of the network
 effect the capacity, that is why the suggested approach can be used if one plans to build or modernize 
 the network. Indeed, it is easy to calculate the capacity for different versions of the net 
 in order to find the optimal. 
 
 \section{The entropy efficiency}

 Let a node $\omega$ from the network $\Omega$ uses files from $F = \{f\}$ with a certain  probability distribution.  
 Our goal is to describe a so-called entropy efficiency which is a  measure of the performance for this case. 
  In order to model this situation 
  we suppose that  for any node $\omega$ a sequence of read files $f_1, f_2, ... , $ 
  is generated by stationary and ergodic process (note, that nowadays this model is one of the most general and  popular).   
  Again, we use ideas and concepts of Information Theory and first give some required  definitions.
  
Let there be a stationary and ergodic process $ z $  generating letters from a finite alphabet $A$ 
(the definition of stationary 
ergodic process can be found, for ex., in \cite{co}).
The $n-$order Shannon entropy and the limit Shannon entropy are defined as follows:
\begin{equation}\label{ent}
 h_n(z) = -  \frac{1}{n+1}\sum_{u \in A^{n+1} } P_z(u) \log P_z(u) ,   $$ 
$$h_\infty(z) = \lim_{n \rightarrow \infty} h_n(z)
\end{equation}
where 
$n \ge 0$ , $P_z(u)$ is the probability that $z_1 z_2 ... z_{|u|} $ $ = u$ (this limit always exists, see  \cite{co,S}).
We will also consider so-called  i.i.d. sources. By definition, they generate independent and identically 
distributed random variables from some set.

Now we can define the entropy efficiency.  
\begin{definition}  
 Let there be a network $\Omega$ and for 
 node $\omega \in \Omega$  a sequence of read files $f_1, f_2, ... , $  is generated by a stationary ergodic source
 $\varphi_\omega$ and
 let $\tau_\omega(f'')$ be the  time of reading the file $f''$ by $\omega$.
 Then  {\it the entropy efficiency } is defined as follows:
\begin{equation}\label{eff}
c(\omega,\varphi_\omega) = h_\infty(\hat{f}) / \sum_{f \in F} P_{\varphi_\omega}(f) \tau_\omega(f) ,
\end{equation}
where $ P_{\varphi_\omega}(f'')$ is the probability that $f_1=f'' $.
The entropy efficiency of the network $\Omega$ equals
$$
c(\Omega, \hat{\varphi}) = \sum_{\omega \in \Omega} c(\omega,\varphi_\omega) \,
$$
where $\hat{\varphi} = ( \varphi_{\omega_1}, ... ,  \varphi_{\omega_{|\Omega|}} )$.
\end{definition}

 Informally, the Shannon entropy is a quantity of information (per letter),
 which can be transmitted and the denominator in (\ref{eff}) is the average time  of reading of one file.   
% It will be shown at this part that  the capacity is a maximal performance which can be achieved for a certain probability distribution read files. 

The entropy efficiency can be estimated basing on statistical estimations of 
frequencies and it gives a possibility to investigate  the influence of different
 parameters on the entropy efficiency.  

It is easy to see that the entropy efficiency  (\ref{eff})  is maximal, if 
the sequence  of files $ f_1 f_2 ... $ 
is generated by an i.i.d. source with probabilities  
$p^*(f)   = X_0^{- \tau_{\omega(f)}(f)}, $    where $X_0$  is the largest real solution to 
the equation (\ref{X}), $f \in F$. Indeed, 
having taken into account that $h_\infty(\omega) = h_0(\omega)$ for i.i.d. source \cite{co} and  the definition of entropy
(\ref{ent}),  the direct calculation of $c(\omega,p^*) $ 
in (\ref{eff}) shows that $c(\omega,p^*) = $ $ \log X_0$ and, hence, 
$c(\omega,p^*)$ equals the capacity $ C(\omega).$

It will be convenient to combine  the results about the  capacity 
and the entropy efficiency in the following statement: 

\begin{claim}\label{TR}
Let  there be a a network $\Omega$ with a set of files $F$, a node $\omega$
 and let $\tau_\omega(f)$ be the  time of reading of the file  $ f$.
  Then 
 the following equalities are valid:
\begin{itemize}
 \item[i)]   The  capacity $C(\omega)$   (\ref{cap}) equals $\log X_0,$ 
where $X_0$  is the largest real solution to 
the equation (\ref{X}). 
 \item[ii)] The entropy efficiency  (\ref{eff})  is maximal if the reading files are generated by an i.i.d. source 
with probabilities  
$p^*(f) = X_0^{- \tau_\omega(f)},$ $f \in F$.
\end{itemize}
\end{claim}

So, we can see that the node capacity is the maximal value of the entropy efficiency, and this maximum can be  attained
 for a certain file distribution. This fact can be useful for applications. 
    For example, if the entropy efficiency is much less than the  capacity, it could mean that  a different
    scheme of caching should be used.
    In other words, information about the network capacity and the entropy efficiency can be
 useful for network designers and manufacturers.

\section*{Acknowledgment}
% optional entry into table of contents (if used)
%\addcontentsline{toc}{section}{Acknowledgment}
Research  was supported  by  Russian Foundation for Basic Research
(grant no. 12-07-00125).

\end{document}